
\input epsf
\input amssym.def 

\magnification= \magstep1
\tolerance=1600 
\parskip=5pt 
\baselineskip= 6 true mm \mathsurround=1pt

\font\smallrm=cmr8

\def\secbreak{\vskip12pt plus 1in \penalty-200\vskip 0pt plus -1in} 

\def\Narrower{\par\narrower\noindent}   
\def\Endnarrower{\par\leftskip=0pt \rightskip=0pt} 
                
\def\a{\alpha}          \def\b{\beta}     
\def\d{\delta}            \def\e{\varepsilon}

             \def\j{\psi}    
         \def\s{\sigma}

  \def\OO{{\cal O}}

\def\cl{\centerline}    
\def\ni{\noindent}                     
                 \def\bra{\langle}       \def\ket{\rangle}
       
\def\fn#1{\ifcase\noteno\def\fnchr{*}\or\def\fnchr{\dagger}\or\def
        \fnchr{\ddagger}\or\def\fnchr{\rm\S}\or\def\fnchr{\|}\or\def
        \fnchr{\rm\P}\fi\footnote{$^{\fnchr}$} {\scrunch#1\toe} 
        \global\advance\noteno by 1\ifnum\noteno>5\global\advance\noteno by-6\fi}
        \def\scrunch{\baselineskip=10 pt \smallrm}
        \def\toe{\vphantom{$p_\big($}}
        \newcount\noteno
      
\def\ddt{{{\rm d}\over {\rm d}t}}
\def\ffract#1#2{{\textstyle{#1\over#2}}}
\def\fract#1#2{\raise .35 em\hbox{$\scriptstyle#1$}\kern-.25em/
        \kern-.2em\lower .22 em \hbox{$\scriptstyle#2$}}

\def\half{\ffract12} 

\def\part#1#2{{\partial#1\over\partial#2}} 
 \def\ref#1{${\vphantom{)}}^#1$}

\def\bbf#1{\setbox0=\hbox{$#1$} \kern-.025em\copy0\kern-\wd0
        \kern.05em\copy0\kern-\wd0 \kern-.025em\raise.0433em\box0}              

\def\ref#1{$^{#1}$}
\def\ra{\rightarrow}

\vglue 1truecm
\rightline{THU-96/39}
\rightline{quant-ph/9612018}
\bigskip
\cl{\bf QUANTUMMECHANICAL BEHAVIOUR IN A}\smallskip
\cl{\bf DETERMINISTIC MODEL}

\bigskip

\cl{G. 't Hooft }
\bigskip
\cl{Institute for Theoretical Physics}
\cl{University of Utrecht, P.O.Box 80 006}
\cl{3508 TA Utrecht, the Netherlands}
\smallskip\cl{e-mail: g.thooft@fys.ruu.nl}
\bigskip\hrule\bigskip
\ni{\bf Abstract} 
\Narrower A deterministic model with a large number of continuous and
discrete degrees of freedom is described, and a statistical treatment
is proposed. The model exactly obeys a Schr\"odinger equation, which
has to be interpreted exactly according to the Copenhagen
prescriptions. After applying a Hartree-Fock approximation, the model
appears to describe genuine quantum particles that could be used as a
starting point for field variables in a quantum field theory. In the
deterministic model it is essential that information loss occurs, but
the corresponding quantum system is unitary and exactly preserves
information. \Endnarrower

\bigskip\medskip\hrule\secbreak

This paper is intended to exhibit the mathematical nature of a model
that could be used for further study of the quantum mechanical nature
of this world, but the philosophical implications will not be
addressed, nor will we discuss possible implications in connection with
the usual quantummechanics paradoxes such as the Einstein Rosen
Podolski paradox~-- these are postponed to a more elaborate publication.
In the present paper, the reader is invited to draw his or her own
conclusions concerning the relevance of this model.
 
First, consider one continuous parameter $q(t)\in\Bbb R$ (boundary conditions
on $q$ will be postponed to later) and one
discrete parameter \hbox{$s =\pm 1$}, and let them obey the (deterministic)
time evolution equations
$$\ddt q(t)=s \ ;\qquad\ddt s =0\,.\eqno(1)$$
This means that we have a `particle' moving along the real line
with velocity either $+1$ or $-1$.
Defining the Hamiltonian
$$H=s\,p\,,\eqno(2)$$
allows us to write the Hamilton equations
$$\ddt q=\{q,\,H\}\ ;\qquad\ddt s =\{s ,\,H\}\ ;\qquad
\hbox{if}\qquad\{q,\,p\}=1\,,\quad\{s,\,p\}=\{s,\,q\}=0\,.\eqno(3)$$

Of course, if we start with a probability distribution $P(q,s ,0)$ at $t=0$,
we have as a solution
$$P(q,s ,t)=P(q-s\,t,s ,0)\,,\eqno(4)$$
and one can write $P\equiv|\j|^2$, where $\j$ is a wave function, also satisfying
$$\j(q,s ,t)=\j(q-s\,t,s ,0)\,.\eqno(5)$$
This $\j$ obeys the Schr\"odinger equation\ref1
$$\ddt\j(q,s ,t)\,=\,-i\hat H\j(q,s ,t)\ ,\qquad\hbox{if}\qquad 
\hat H=s\,\hat p\ ,\quad\hat p=-i\part{ }q 
\ .\eqno(6)$$
Writing things this way, we see that the `quantum' model of Eq.~6 is physically
and mathematically identical to the classical one. Here, we write `quantum'
between quotation marks because it can hardly be called a quantum
theory in the usual sense: the wave function $\j$ does not spread as
time proceeds (so one cannot envisage interference experiments), and the
Hamiltonian $\hat H$ is not bounded from below, so that there is no
ground state. It is of importance, however, to emphasize that it is
entirely legal to attach to this wave function $\j$ a Copenhagen
probability interpretation. Quantum mechanical superposition of states
$\j$ is permitted in the usual way. Later, we will see how a lower bound in the
Hamiltonian of a true quantum theory may arise. A lower bound will be necessary 
since we wish to have a ground state, or vacuum state.

Next, let us take two sets of such variables, $q_1$, $q_2$, $s_1$ and
$s_2$. Again, $q_i$ are real numbers and $s_i$ are $\pm1$. 
If the Hamiltonian is taken to be (from now on we omit the hats)
$$H_0=s_1 p_1+s_2 p_2\ ,\qquad[q_i,\,p_j]=i\d_{ij}\ ;\quad[s_i,\,s_j]=0\ ,
\eqno(7)$$
then we have two (distinguishable) particles moving as before.

Now, we introduce as interaction 
$$s_1\ra-s_1 \qquad \hbox{if}\qquad  q_2=c\ ,$$
\ni where $c$ is a fixed number. This means that particle~1 flips its
velocity whenever particle~2 crosses the point $c$. Clearly, this is a
deterministic law.  We can express this interaction by means of an
extra term in the quantum Hamiltonian:
$$H=H_0+H_1\ ;\qquad H_1=\half\pi(\s_1^1-1)\,\d(q_2-c)\,,\eqno(8)$$
where we introduced the Pauli matrices
$$\s_i^3\equiv s_i\ ,\qquad [\s^a_i,\,\s^b_j]=2i\,\d_{ij}\,\e^{abc}\,
\s_i^c\ .\eqno(9)$$ 
This Hamiltonian works because $q_2$ moves with a fixed velocity; the factor
$\half\pi$ is exactly what is needed to flip $\s_1^3$ over. The $-1$ in
Eq.~8 is there to ensure that the effect of this Hamiltonian is a flip
{\it without\/} a sign change in the wave function:
$$\exp\big(-\half i\pi(\s_1^1-1)\big) = 
\s_1=\pmatrix{0&1\cr 1&0}\,.\eqno(10)$$
The positive signs here are important because we wish to find solutions
for $\j$ that are either positive real numbers or at most have slowly
varying phase factors, not depending very much on $s_i$. This will
allow us later to construct approximate solutions. The {\it overall
sign\/} in the Hamiltonian $H_1$ is at first sight arbitrary, but this
choice turns out to be important later (see remarks near the end of
this paper).

Next, it will be of essential importance to introduce {\it information
loss\/} at the deterministic level: states that are initially different
may evolve into the same final state. This is what will make our model
truly quantum mechanical. The resulting quantum theory will still be
time-reversible and unitary, because the physical states will be
defined to be only those that can be reached in the infinite future\ref2.
When $q_2$ passes a point $a$, the discrete variable $s_1$ will jump to
$+1$ if it was $-1$ before, and if it was $+1$ it will stay $+1$.
This is described by the matrix
$$\pmatrix{1&1\cr 0&0}=\half(i\s^2+\s^3+\s^1+1)\,.\eqno(11)$$
Since
$$\exp\big(i\a\,\s^2+\a(\s^3+\s^1-1)\big)=\pmatrix{1&1-e^{-2\a}\cr
0&e^{-2\a}}\,, \eqno(12)$$ we can use the interaction Hamiltonian
$$H_2=\a\big(-\s^2+i(\s^1+\s^3-1)\big)\,\d(q_2-a)\,,\eqno(13)$$
where $\a$ is big, but not too big for our later approximation scheme.
Note that $H_2$ by itself has one real Eigenvalue $0$ with Eigenvector
$|\j(1)\ket=\left({1\atop0}\right)$ and a complex one,
\hbox{$-2i\a\d(q_2-a)$,} with Eigenvector
$|\j(2)\ket={1\over\sqrt2}\big({1\atop -1}\big)$. In the total
Hamiltonian, at a later stage, we will keep only the real Eigenvalues.
Similarly, the interaction Hamiltonian
$$H_3=\b\big(\s^2+i(\s^1-\s^3-1)\big)\,\d(q_2-b)\,,\eqno(14)$$
with sufficiently large $\b$, will turn the spin $s_1$ to $-1$ if
$q_2$ passes the point $b$. Again, additive constants in the Hamiltonians
$H_2$ and $H_3$ have been chosen such that the transition matrices such 
as Eq.~(11) carry positive signs only.

So-far, these Hamiltonians do not look useful for simulating more
interesting quantum systems such as harmonic oscillators. But now we
concentrate on the many-particle case.\fn{In a more realistic theory, at
a later stage, we may be interested in regarding the $q_i$ as field
values, and replace the index $i$ by a 3-space coordinate.} Let there
be $N$ continuous degrees of freedom $q_i$, $i=1,\,\dots,\,N$, and $N$
discrete operators $s_i=\s_i^3$. Let $p_i$ and $\s^a_i$ obey the
commutation rules~(7) and~(9). The Hamiltonian is generalized into
$$\eqalign{H=&\sum_i\s_i^3p_i\,+\sum_{i,j,k}\Big(\half\pi(\s^1_i-1)\,
\d(q_j-c_{ij}^k) +\a\big(-\s_i^2+i(\s_i^1+\s_i^3-1)\big)\d(q_j-
a_{ij}^k)\ +\cr &+\b\big(\s_i^2 +i(\s_i^1-\s_i^3-1)\big)
\d(q_j-b_{ij}^k)\Big)\,,\cr}\eqno(15)$$
The big summation here goes over values of $j\ne i$, and at each pair
$(i,j)$ there may be several $k$ values.

The points $a_{ij}^k,\ b_{ij}^k$ and $c_{ij}^k$ are the free parameters of
the model. As must be clear from the foregoing, this model is entirely
deterministic; when $q_j$ takes one of the values $c_{ij}^k$, the velocity
of the coordinate $q_i$  changes sign, if $q_j=a_{ij}^k$, this velocity
becomes $+1$, if $q_j=b_{ij}^k$, this velocity becomes $-1$.

We are particularly interested in the case where, for any given $j$, the
points $c_{ij}^k$ are fairly densely and smoothly distributed (the points
$a_{ij}^k$ and $b_{ij}^k$, where information loss takes place, may be very 
scarce.) We imagine that the
deterministic model will become chaotic. In order to approximate its
behaviour statistically, we construct an approximate solution of the
Eigenvalues and Eigenstates of $H$, using a  `Hartree-Fock' approximation.
Take a trial wave function $\j$  of the form
$$\j\big(\{q_i,s_i\}\big)=\prod_i\j_i(q_i,s_i)\,,\eqno(16)$$
where $\j_i$ are sufficiently smooth single-particle wave functions.

Standard variation techniques tell us that the best functions $\j_i$ are
obtained as follows.  Define
$$\eqalignno{\half\pi\sum_{j,k}\bra\j_j|\d(q_j-c_{ij}^k)|\j_j\ket&\ 
=\ C_i\ ;&(17)\cr
\a\sum_{j,k}\bra\j_j|\d(q_j-a_{ij}^k)|\j_j\ket&\ =\ A_i\ ;&(18)\cr
\b\sum_{j,k}\bra\j_j|\d(q_j-b_{ij}^k)|\j_j\ket&\ =\ B_i\ ;&(19)\cr
\sum_{i,k}\bra\j_i\big|\half\pi\d(q_j-c_{ij}^k)\,(\s^1_i-1)+\a\big(-\s^2_i&
+i(\s^1_i+\s^3_i-1)\big)\d(q_j-a_{ij}^k)\  +&{  }\cr +\,\b\big(\s^2_i
+i(\s^1_i-\s^3_i-1)\big)\d(q_j-&b_{ij}^k)\big|\j_i\ket \ =\ V_j(q_j)+iW_j(q_j)\,.&(20)\cr
 }$$

The one-particle wave functions are then seen to obey 
$$\eqalign{H_i\j_i= E_i\j_i&\ ;\qquad\qquad H_i\ =\ \s^1_i(C_i+iA_i+iB_i)\ +\cr
+\ \s^2_i(B_i-A_i)+\s^3_i&(p_i+iA_i-iB_i)-C_i-iA_i-iB_i
+V_i(q_i)+iW_i(q_i)\,.\cr}\eqno(21)$$

If we are only interested in the Eigenvalues $E_i$, a renormalization of the
wave functions allows us to replace $p_i$ by 
$$\hat p_i\equiv p_i+iA_i-iB_i\,.\eqno(22)$$
The Eigenvalues $E_i$ are (for simplicity we drop the index $i$):
$$\eqalign{E=&-C-i(A+B)+V(q)+iW(q)\pm\sqrt{\big(C+i(A+B)\big)^2+(B-A)^2+\hat p^2}\,=\cr
=&-C-i(A+B)+V+iW\pm\left[C+i(A+B)+{(B-A)^2+\hat p^2\over 2\big(C+i(A+B)\big)}
+\OO(\hat p^4)\right]\,.\cr}
\eqno(23)$$

\midinsert\epsffile{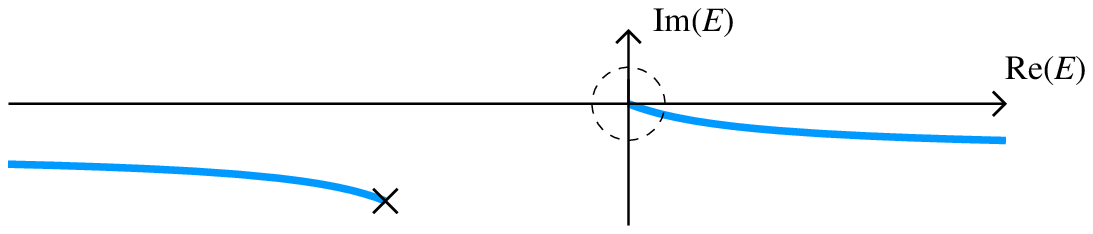}
\cl{\scrunch Fig. 1. The complex energy espectrum. Physical states will 
mostly be within the dotted circle.}
\endinsert

Now, if we had no information loss, that is, $A=B=W=0$, we would have
obtained a quantum particle with mass $\pm C$. The negative mass values
are a serious problem here. In deterministic models, the negative
Eigenvalues of the Hamiltonian have always been a serious difficulty.\ref{1,\,3}
The most important result we wish to report here is the effect of
information dissipation. We still assume $A$ and $B$ to be very small
compared to $C$.  In this case, the Eigenvalues with positive signs in
Eq.~(23) are practically real, whereas the others have substantial
negative imaginary parts. These must be the states that do not survive
at $t\ra\infty$.  The other states have a tiny negative imaginary part.
Knowing that there must exist many real Eigenvalues, we imagine that
higher order corrections will be big enough to remove the tiny imaginary
parts for many of the lower energy states (inside the dotted circle in
Figure~1.

We find that only the states with positive energies survive. This sign
preference may be explained as follows. If we had chosen the signs in
Eq.~(10) oppositely, we would have found that only the negative energy
values survive. Thus, the sign choice in Eq.~(10) must be seen as
establishing the sign convention for the entire Hamiltonian.
 
Corrections to the Hartree-Fock aproximation will give interactions
between the particle-like objects. Since their matrix elements are
non-vanishing, there will be contributions from virtual negative-energy
states, but these will not occur in the final states because they are
non-physical. By construction, the exact evolution matrix $U=e^{-iHt}$
will be unitary within the sector of the physical states, which is why
we expect the exact physical spectrum to contain real $E$ values only.

Substituting the solutions for $\j_i$ back into Eq.~20, we find that
$V$ and $W$ vanish as soon as $B=A$, since then $\j_i(q_i,s_i)=1$. But
it is easy to introduce non-trivial boundary conditions for the $q$'s,
so that $\bra p_i^2\ket\ne0$, and $\bra\s^1_i\ket\ne0$, in which case 
non-trivial potentials $V$ and $W$ may emerge (with $|W|\ll|V|$).
Here we observe the effects of the $-1$ in Eq.~8: it removes unnecessary
large interaction terms so that the Hartree Fock approximation works
optimally. There exist of course more ways to generalize the model.
Most essential features seem te be the fact that there are {\it continuous\/}
variables $q_i$ (so that an infinitesimal time shift can be defined),
and {\it discrete\/} variables $s_i$ (allowing us to introduce information
loss).

Our arguments claiming that non-Hartree Fock corrections will produce a
real (and positive) energy spectrum are as yet delicate, and may be
called inconclusive.  We suggest that whether or not a non-trivial
spectrum of real energy values emerges may depend on further details of
the model. It will probably be essential that our model be chaotic, a
feature that will depend on details in a complicated way. An other way
to look at the model is to consider the $N\ra\infty$ limit, and assume
that every `particle' interacts with every other one.\fn{See footnote
on Page 4} In this limit, $C$ may tend to infinity and if we keep $A$
and $B$ small, the positive energy values will become real. We have a
real quantum theory. Needless to say, the model can be generalized in
many ways. Here, we merely wanted to point out some of its most
essential features.

\secbreak
{\ni \bf References}\medskip
\item{1.} G. 't Hooft, {\it J. Stat. Physics\/} {\bf 53 (1988)} 323; 
{\it Nucl. Phys.\/} {\bf B342} (1990) 471.
\item{2.} G. 't Hooft, to be publ.
\item{3.} G. 't Hooft, K. Isler and S. Kalitzin, {\it Nucl. Phys.}
{\bf B386} (1992) 495.
 
\bye